\documentclass[a4paper]{article}
\usepackage[T1]{fontenc}
\usepackage{authblk}
\usepackage{graphicx}

\title{Piggyback resistive Micromegas}
\author[1]{D.~ Atti\'e}
\author[1]{A. Chaus}
\author[1]{D. Durand}
\author[1]{D. Deforges}  
\author[1]{E. Ferrer-Ribas\footnote{Corresponding author.}}
\author[1]{J. Gal\'an}
\author[1]{Y. Giomataris} 
\author[1]{A. Gongadze\footnote{Present address: Dubna}}
\author[1]{F. J. Iguaz\footnote{Present address: Instituto de F{\'i}sica Nuclear y Altas Energ{\'i}as, Universidad de Zaragoza, Zaragoza, Spain.}}
\author[1]{F. Jeanneau}  
\author[2]{R. de Oliveira} 
\author[1]{T. Papaevangelou}  
\author[1]{A. Peyaud} 
\author[2]{A. Teixeira}

\affil[1]{IRFU, Centre d'\'Etudes Nucl\'eaires de Saclay (CEA-Saclay), Gif-sur-Yvette, France}
\affil[2]{European Organization for Nuclear Research (CERN), Gen\`eve, Switzerland}

\begin{document}
\maketitle
\abstract{Piggyback Micromegas consists in a novel readout architecture where the anode element is made of a resistive layer on a ceramic substrate. The resistive layer is deposited on the thin ceramic substrate by an industrial process which provides large dynamic range of resistivity (10$^6$ to 10$^{10}$\,M$\Omega$/square). The particularity of this new structure is that the active part is entirely dissociated from the read-out element. This gives a large flexibility on the design of the anode structure and the readout scheme. Without significant loss, signals are transmitted by capacitive coupling to the read-out pads. The detector provides high gas gain, good energy resolution and the resistive layer assures spark protection for the electronics. This assembly could be combined with modern pixel array electronic ASICs. First tests with different Piggyback detectors and configurations will be presented. This structure is adequate for cost effective fabrication and low outgassing detectors. It was designed to perform in sealed mode and its long term stability has been extensively studied. In addition perspectives on the future developments will be evoked.}

\section{Introduction}\label{sec:intro}
Gaseous detectors suffer from the secondary effects of discharges induced by heavily ionizing particles producing large deposits. Even if these discharges are not destructive, as in the case of Micromegas detectors \cite{mm96,mm98}, they can affect the detector behaviour limiting its operation. The most common effects are: i) introduction of dead time due to the
required power supply recovery time; ii) reduction of  the mean life time of the detector by the heating of the surrounding materials around the affected areas and finally  iii) increase of the risk of damaging the readout electronics due to the high currents reached by these spark processes. Different approaches exist to diminish these secondary effects. The solution chosen for the upgrade of the forward ATLAS muon system in order to cope with the high luminosity of the LHC (HL-LHC) project is to use resistive strips on top of the anode plane \cite{ResistiveMAMMA}. This solution was first introduced in \cite{Dixit} with a resistive foil in order to obtain charge dispersion of the signal in the context of the TPC for the ILC. In the case of the Micromegas coupled to a Medepix chip, protection is absolutely mandatory as discharges can melt or evaporate the surface of the silicon chip. In order to protect the chip from discharges  a high resistive layer of amorphous silicon is deposited on the chip.
Here  we recall  the {\it Piggyback} concept where a Micromegas mesh is  layed on a thin resistive layer deposited on an adequate insulator.  This  new approach is described in detail in \cite{piggyback}. In this paper, section 2 describes briefly the principle and experimental set-up. Section 3 is devoted to characterization measurements including gain and energy resolution, rate capability, operation with a Medipix chip and novel first results on sealed mode operation. Finally  conclusions and perspectives are discussed in section 4.

\section{Principle and experimental set-up }
\subsection{Principle}
\begin{figure}[tbp] 
\centering
\includegraphics[width=0.6\textwidth]{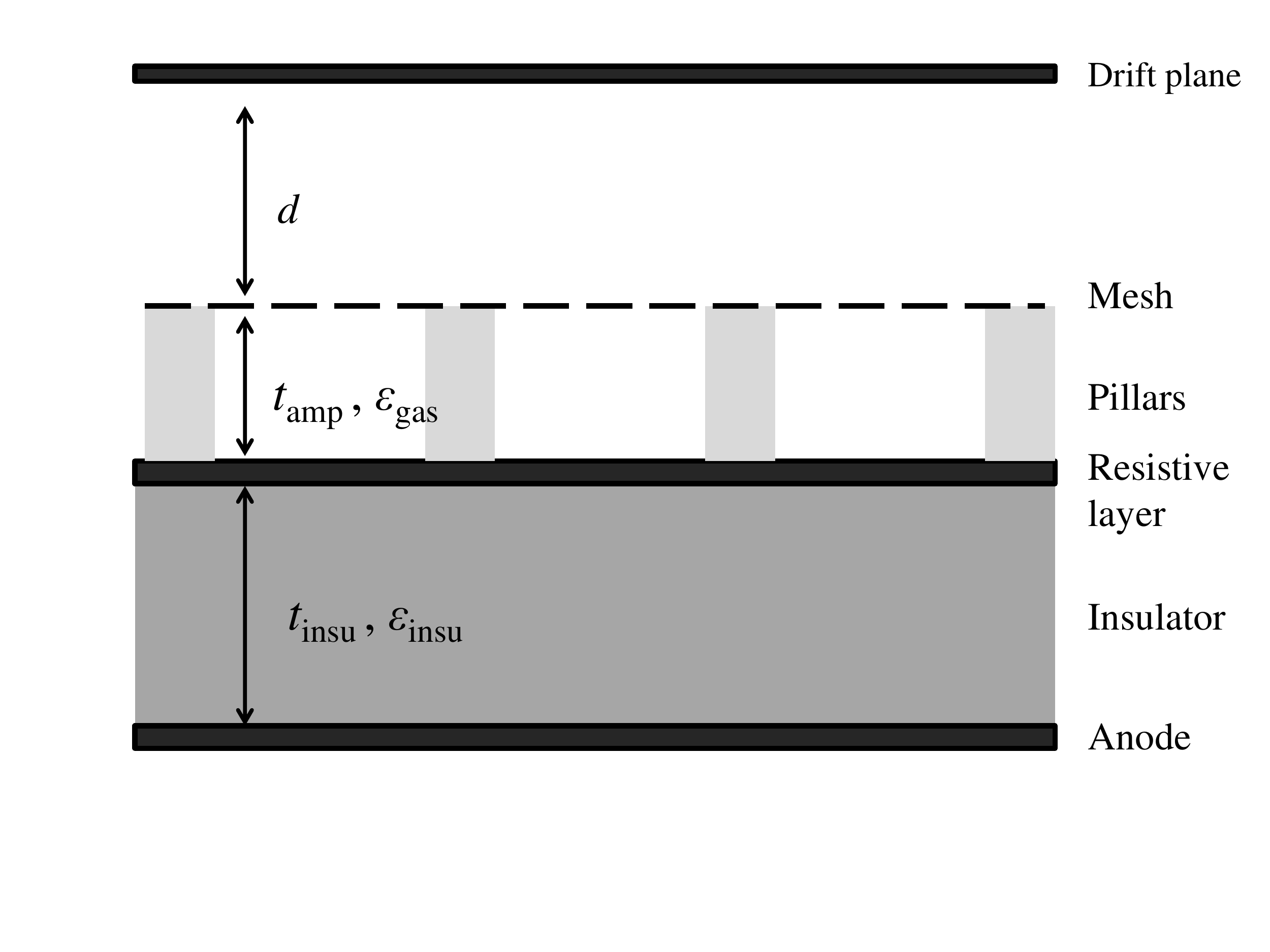}
\caption{Sketch of the Piggyback structure.}
\label{fig:principle}
\end{figure}

The concept is based on the idea of separating the detector active part from the readout plane. The signal is transmitted by capacitive coupling on the readout pads. This idea was inspired from an analogous development on Parallel Plate Avalanche Counter \cite{PPAC}. The operation principle is sketched in figure~\ref{fig:principle}. The standard elements of a Micromegas detectors are shown: the drift plane, the amplification gap defined by a woven mesh connected to high voltage laying on pillars of around 100\,$\mu$m height and a resitive layer. The novelty is the fact that the resistive layer is deposited on a insulator substrate that will be posed on the readout plane. The Micromegas detector operates as usual in the proportional avalanche mode inducing signals on the resistive anode plane. The structure needs to   be optimised in such a way that the electronic signal is not lost through the resistive layer but is propagated to the readout plane. In practice the optimisation consists in keeping the insulator thickness small compared to the product of the amplication gap and the dielectric constants ratio in the following way:
\begin{equation}
t_{insu}<<t_{amp}\frac{\epsilon_{insu}}{\epsilon_{gas}}
\end{equation}
where $\epsilon_{insu}$ is the dielectric constract of the insulator, $\epsilon_{gas}$ is the gas dielectric constant and $t_{insu}$ and $t_{amp}$ being the thickness of the insulator layer and of the amplification gap respectively.

Materials with large dielectric constants are favoured ($\epsilon_{insu}\gg$10). The first experimental tests have been performed with a bulk micromegas with an  amplification gap, $ t_{amp}$,  of 128\, $\mu$m and $t_{insu}$ of 300\,$\mu$m with  ceramic alumina. For the resistive layer, ruthenium oxide (RuO$_2$) has been chosen for its robustness, stability and wide range of resistivity values.
\subsection{Experimental set-up}
The first tests were performed with  bulk detectors of  a 3$\times$3\,cm$^2$ active zone and the characteristics described above. The conversion gap was 1 cm defined by a mesh frame above the amplification gap. A copper layer covering the whole surface of the detector was placed at the back of the ceramic as one single anode pad. The mesh voltage was varied from
300 to 450\,V, the drift from 400 to 1200\,V  the anode was connected to ground. Mesh and drift voltages were independently powered by a CAEN N471A module. The avalanche in the
Micromegas structure induces a negative signal in the mesh and a positive one in the anode. Both
signals were respectively monitored by two CANBERRA preamplifiers to evaluate their efficiency.
The preamplifier outputs were fed into two CANBERRA 2006 amplifiers and subsequently into a
multichannel analyzer AMPTEK MCA-8000A for spectra acquisition. Each spectrum was fitted
by two Gaussian functions, corresponding to the K$_\alpha$ (5.9\,keV) and K$_\beta$ (6.4\,keV) lines of the $^{55}$Fe
source; the mean position and width of the main peak were obtained for each fit. Two types of gas mixtures were used in these tests: 
Argon + 5\%  Isobutane and Neon + 5\%  Ethane.

\section{Characterisation}
%
\subsection{Gain and energy resolution}
\begin{figure}[tbp] 
\centering
\includegraphics[width=1\textwidth]{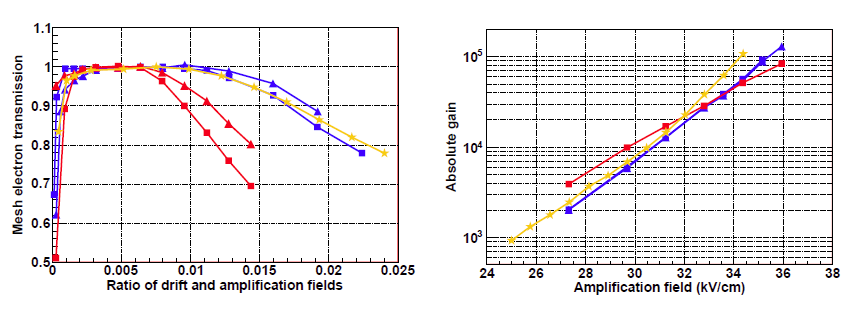}
\caption{Left: Electron transmission as a function of drift and amplification ratio. Right : Gain as a function of amplification field. The blue curves show results with Argon + 5\% Isobutane and red with Neon + 5\% Ethane. The orange stars show the behaviour with a standard bulk (128 $\mu$m thickness gap)  detector in Argon + 5\% Isobutane for comparison.}
\label{fig:first_test}
\end{figure}

\begin{figure}[tbp] 
\centering
\includegraphics[width=1\textwidth]{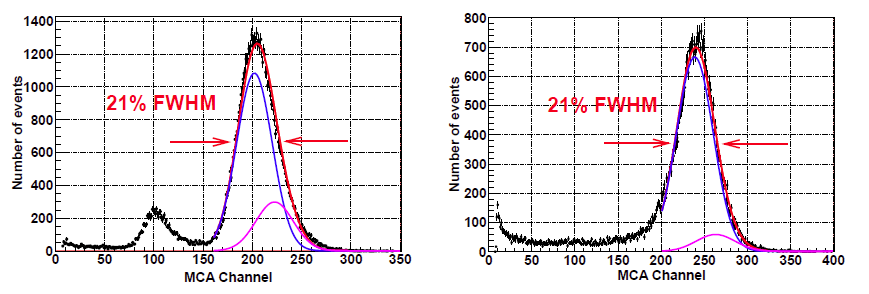}
\caption{  $^{55}$Fe spectra obtained with Argon + 5\%  Isobutane (left) and Neon + 5\%   Ethane (right). }
\label{fig:resolution}
\end{figure}

At a fixed amplification field a scan of the drift field was performed in order to obtain the electron transmission shown in figure~\ref{fig:first_test}. The plot shows the typical plateau of electron transmission for a wide range of low drift fields where the funnel shape of the electric field lines maximises electron transmission. At high fields, the mesh stops being transparent as some of the field lines stop at the mesh and part of the electrons are lost.  The operation point of the detector is chosen in the plateau  and the  gain of the detector was studied by varying the amplification field. The gain of the detector is shown for  the two tested gases in figure~\ref{fig:first_test}. The two Piggyback detectors show slightly lower gains than a standard 128\, $\mu$m bulk detector but however reaching gain values up to 10$^5$. The energy resolution of  21\,\% (FWHM)  at 5.9 keV is shown in figure~\ref{fig:resolution}.

\subsection{Rate capability}
\begin{figure}[tbp] 
\centering
\includegraphics[width=1\textwidth]{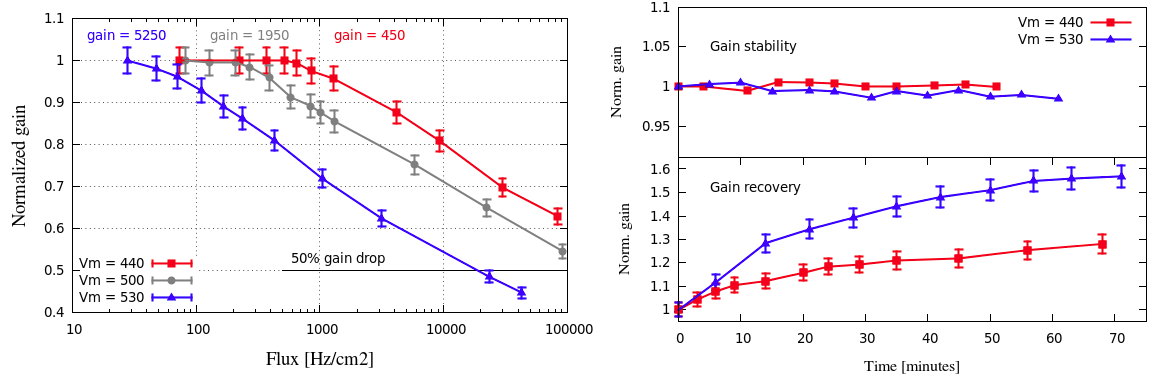}
\caption{Relative gain drop for different X-ray fluxes (left) and amplification gains (right top).  Gain recovery after high flux exposure to low flux transition (right bottom).}
\label{fig:ratecapability}
\end{figure}

The rate capability of the Piggyback detectors was studied using an X-ray generator.  The 8\,keV Copper peak was used as a gain reference. Gain measurements were performed at different X-ray fluxes up to 100\,kHz$/$cm$^2$ shown in figure~\ref{fig:ratecapability}. The "plateau" region that is observed depends on the mesh voltage and extends to higher fluxes for lower values of the amplification field. In the same figure (top-right) the stability of the gain is shown when the system is  activated and illuminated with the X-rays, the stability is within 2\%.

In reference \cite{piggyback} we show a comparison of  the experimental gain and the simulated gain drop as a function of the rate showing a good understanding of the general behaviour with the pertinent parameters.

\subsection{Operation coupled with the MEDIPIX chip}
Medipix2 \cite{bib11} and Timepix \cite{bib12} are CMOS chips which have been used as pixelated readout for Micro-Pattern Gaseous Detectors (MPGDs)~\cite{bib6,bib13}.  They consist of a matrix of 256$\times$256 identical square pixels of 55\,$\mu$m side each defining an active area of 14$\times$16\,mm$^2$. In order to protect the chip from sparks that are fatal,  layers of high resistivity material (amorphous silicon or silicon-rich nitride) are usually deposited on CMOS pixel for discharge protection~\cite{bib14}.

\begin{figure}[tbp] 
\centering
\includegraphics[width=0.4\textwidth]{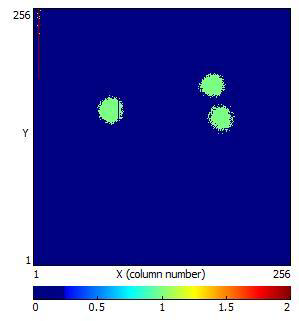}
\caption{Three 5.9 keV photons recorded with a non-protected Medipix2 on top of a Piggyback detector. }
\label{fig:medipix}
\end{figure}

A 30$\times$20\,mm$^2$ piggyback Micromegas detector was built to be used with a non-protected Medipix2/Timepix chips. The amplification gap and the drift height were respectively 128\,$\mu$m and 1\,cm. The ceramic was layed on top a Medipix2 chip without gluing. The detector was tested using  a $^{55}$Fe source. An image of three recorded events is shown in figure~\ref{fig:medipix}. The mesh voltage was 430\,V corresponding to a gain of about 10$^5$  in a gas mixture of argon and 5\% isobutane. The detector was operating under such a high gain for a 24 hours proving the effectiveness of the spark protection scheme. These first results are extremely encouraging, demonstrating the proof of principle of this novel read-out architecture for the Micromegas detector.

\subsection{Operation in sealed mode}
\begin{figure}[tbp] 
\centering
\includegraphics[width=0.4\textwidth]{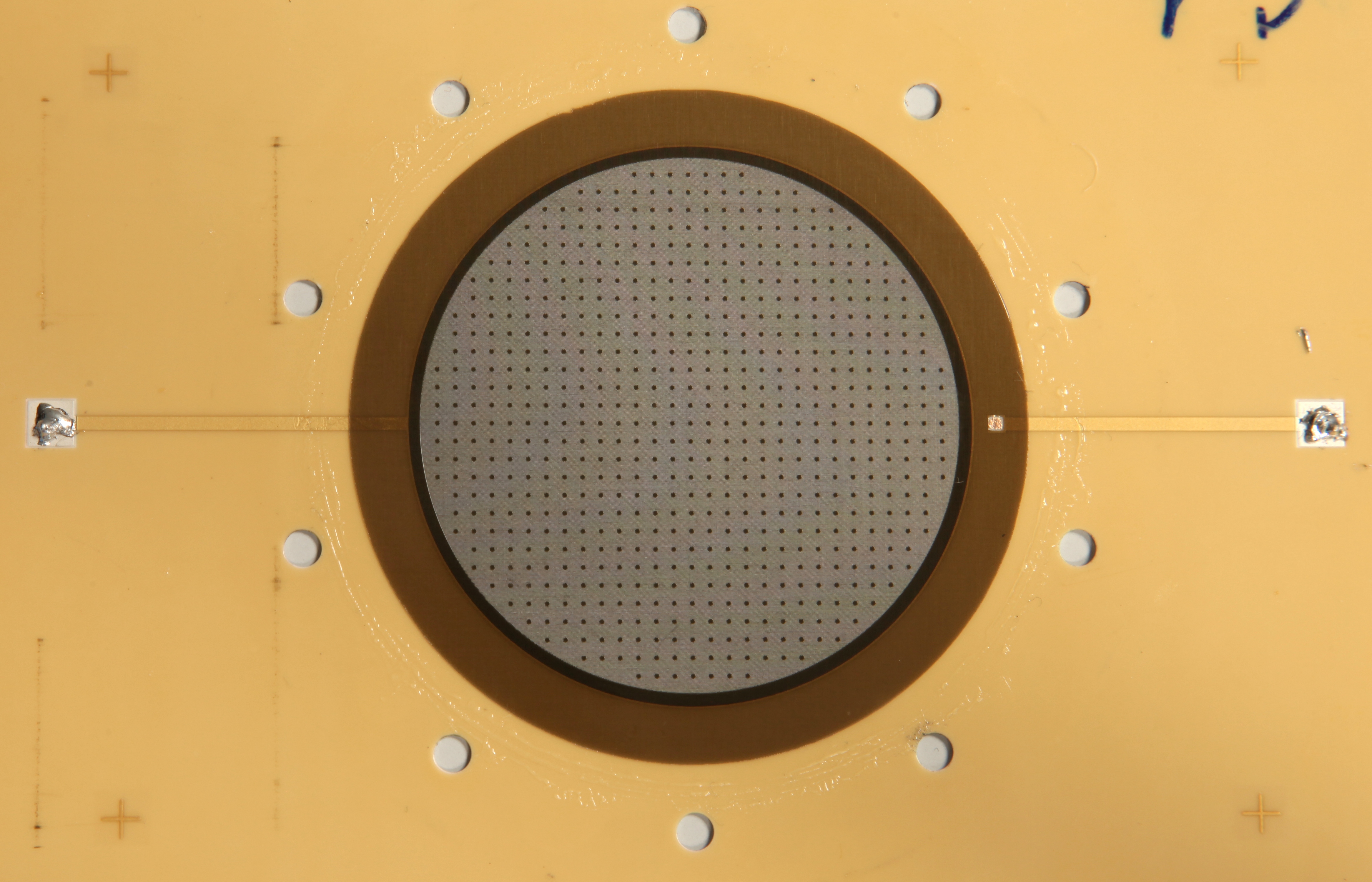}
\caption{Photo of the Piggyback detector used for the tests in sealed mode. The diameter of the active area is of 6\,cm.}
\label{fig:photo}
\end{figure}

\begin{figure}[tbp] 
\centering
\includegraphics[width=0.9\textwidth]{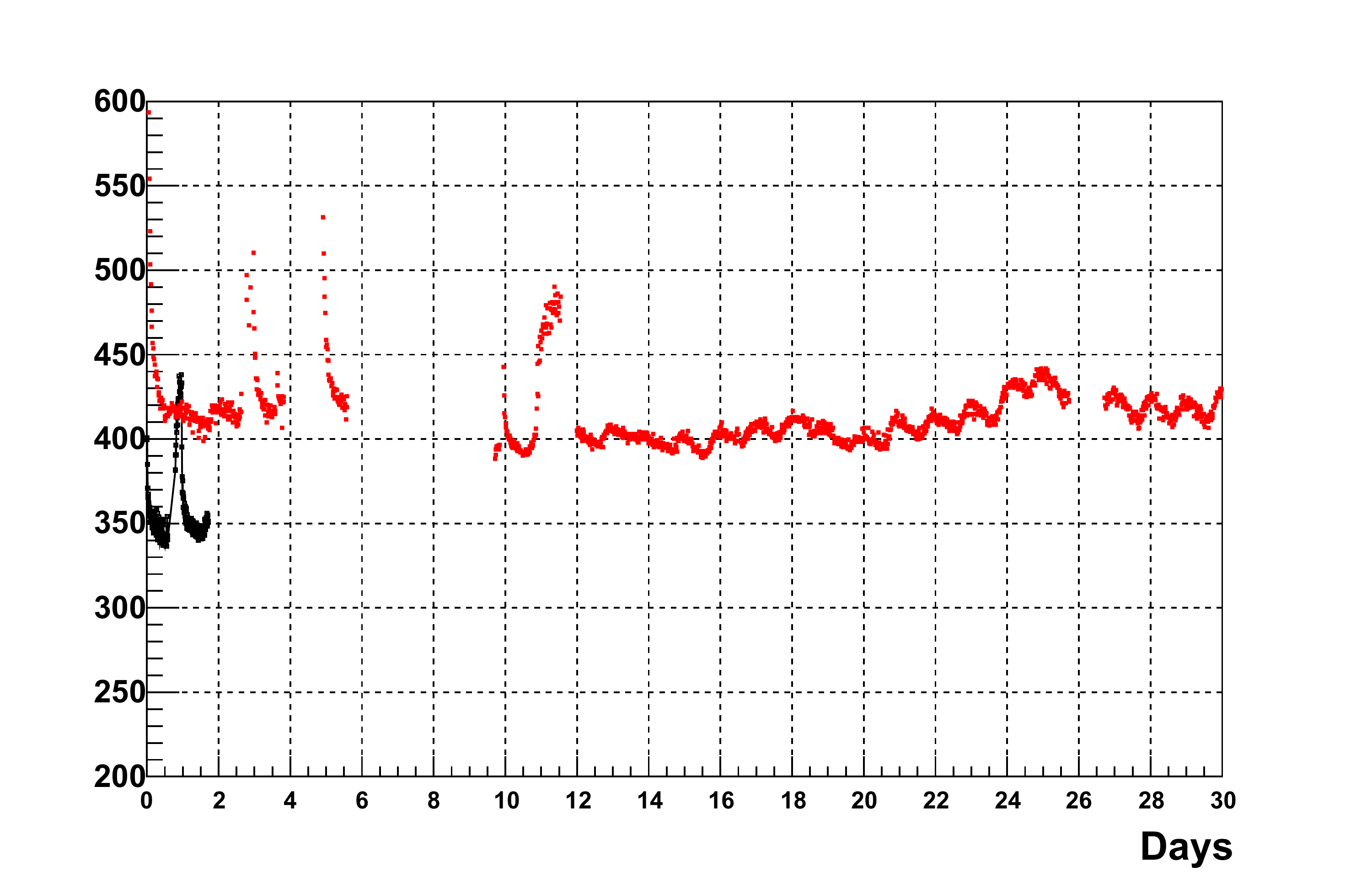}
\caption{Gain in arbitrary units as a function of time. The black points represent data taken with circulating gas during 2 days and red points have been recorded in sealed mode for 30 days with a stability better than 10\%. }
\label{fig:sealedmode}
\end{figure}

A second lot of Piggyback detectors were built increasing the active area to 28\,cm$^2$. A picture of one detector is shown in figure~\ref{fig:photo}.  The detector was tested with 1\, cm conversion gap and in a neon and 10\% ethane mixture. In order to improve leak-tightness of the chamber, the detector chamber was glued to the ceramic layer. The detector was placed over night at 60$^\mathrm{o}$C with a pumping sytem to improve the outgassing. The gas was circulated for 4 hours before isolating  the detector chamber from the gas system by closing the chamber input and output valves. The gas mixture system was shut down. The stability of the detector was monitored using a  $^{55}$Fe source in continous mode (10\,min acquisition every 10\,min) with the AMPTEK MCA for 30 days. The result is shown in figure \ref{fig:sealedmode}. The gain (in arbitrary units) is plotted  as a fonction of days. The black points show the data recorded with the gas circulating while the the red points show the behaviour in sealed mode. It can be observed that the behaviour of the detector in sealed mode was very stable. The mean value of the gain is constant within less than 10\%. The four high spikes observed in the "gas off" curve correspond  exactly to the times that the source was removed and placed back due to charging up effects.The small repetitive fluctuations are probably correlated to day-nights effects. In addition they are probably correlated to the temperature of the laboratory as seen in other sealed gaseous detector~\cite{bellazzini}. This effect will be verified in a future set-up where the pressure and temperature will be monitored. For the basic precautions taken, this result is very encouraging. On the contrary, the day-night effect is more prominent in the "gas on" mode where the behaviour of the detector is more subjected to these changes. The big spike present in the curve is attributed to this effect. We have designed a next set-up where special care has been taken to optimise sealed mode operation. In addition  we will perform systematic cycles of heating and gas flow to reduce outgassing  which should improve the present performance. The gain of the detector will be monitored as long as necessary to be able to measure the gain loss slope.

\section{Conclusion and outlook}
A new type of Micromegas detector has been developped and tested succesfully. We call it Piggyback Micromegas as it can be placed "on the back" of any readout plane with a granularity that can be adapted easily. It consists of a resistive anode deposited by an industrial process on a thin ceramic substrate. The principle is to dissociate the active region from the readout plane providing full spark protection by the resistive layer. This structure provides different advantages, for instance  it allows easier implementation of many integrated chips minimising dead space since
no additional support of interface structure is needed. In addition thanks to the detector materials, ruthenium oxide and ceramics, it  exhibits excellent outgassing properties. This is appropriate for high quality vacuum and it opens the way to  have a fully sealed detector vessel.

Piggyback detectors have been characterised in Argon + 5\%  Isobutane and Ne + 5\%  Ethane mixtures showing a typical behaviour in terms of gain, electron transmission and energy resolution. The Piggyback detector was used with the Medipix2/Timepix chip. The detector was operated at gains of 10$^5$ demonstrating  the effectiveness of the spark protection scheme. These first results are extremely encouraging, prooving the proof of principle of this novel read-out architecture for the Micromegas detector.
The rate capability of the Piggyback detectors has been studied with an X-ray gun and the dependence with resistivity has been studied and understood. We plan to test higher values of resistivity and different thickness of ceramic layer.
We have started to explore the sealed mode operation, in the absence of gas flow. A first test, without any special precaution, showed that the detector can operate in seal mode for at least 30\,days without any significant loss of gain. We will pursue these studies  with an improved set-up and with systematic cycles of heating and gas flow to reduce outgassing where we expect further improvements.


\end{document}